\documentclass[12pt]{article}
\usepackage{amsmath}
\usepackage{graphicx}
\usepackage{amssymb,amsthm}
\usepackage[english]{babel}
\usepackage[textwidth=18.5cm,textheight=22cm]{geometry}

\newcommand{\R}{\mathbb{R}}
\newcommand{\T}{\mathbb{T}}
\newcommand{\U}{\mathbb{U}}
\newcommand{\V}{\mathbb{V}}

\newcommand{\mL}{\mathcal{L}}

\newcommand{\mR}{\mathcal{R}}

\newcommand{\mC}{\mathcal{C}}

\renewcommand{\d}{\mathrm{d}}
\newcommand{\ct}{\boldsymbol{\mathrm{t}}}

\newcommand{\bt}{\boldsymbol{t}}

\newcommand{\be}{\begin{equation}}
\newcommand{\ee}{\end{equation}}
\newtheorem{pro}{Proposition}
\newtheorem{Lemma}{Lemma}
\newtheorem{defi}{Definition}
\newtheorem{teh}{Theorem}

\begin{document}

\title{The double scaling limit method\\ in the Toda hierarchy
\thanks{Partially supported by  MEC
project FIS2005-00319 and ESF programme MISGAM}}
\author{L. Mart\'{\i}nez Alonso$^{1}$ and E. Medina$^{2}$
\\
\emph{$^1$ Departamento de F\'{\i}sica Te\'{o}rica II, Universidad
Complutense}\\ \emph{E28040 Madrid, Spain}\\
\emph{$^2$ Departamento de Matem\'aticas, Universidad de C\'adiz}\\
\emph{E11510 Puerto Real, C\'adiz, Spain} }
\date{} \maketitle
\maketitle \abstract{ Critical points of semiclassical expansions of
solutions to the dispersionful Toda hierarchy are considered and a double scaling
limit method of regularization is formulated.
The analogues of the critical points characterized by
 the strong conditions in the Hermitian matrix model are analyzed and the property of doubling
of equations is proved. A wide family  of  sets of critical points is introduced
and the corresponding double scaling limit expansions are discussed.
 }

\vspace*{.5cm}

\begin{center}\begin{minipage}{12cm}
\emph{Key words:} Toda hierarchy. Double scaling limit.  Hermitian
matrix model.
 \emph{PACS number:} 02.30.Ik.
\end{minipage}
\end{center}
\newpage
\section{Introduction}

In a recent work \cite{mel1} we considered the large $N$ expansion of the Hermitian matrix model
\begin{equation}\label{m1}
Z_N(\bt)=\int \d H \exp\Big(\dfrac{N}{g}\;\mbox{tr}(\sum_{k\geq
1}t_k\,H^k)\Big),\quad H=H^{\dagger},
\end{equation}
from the point of view  of the \emph{dispersionful} Toda hierarchy \cite{tt}
\begin{equation}\label{tod}
\epsilon\,\dfrac{\partial \mL}{\partial t_j}=[(\mL^j)_+,\mL],\quad \mL=\Lambda+u+v\,\Lambda^{-1},\quad \Lambda:=\exp{(\epsilon\,\partial_x)},\quad \epsilon:=g/N,
\end{equation}
where $(\,)_+$ denotes the polynomial part in $\Lambda$. If the equilibrium measure in the large $N$ limit is supported on a single interval (one-cut case), then the partition function   becomes
a tau-function $\tau(x,\bt)$ of this hierarchy at $x=g$. The corresponding solution of \eqref{tod}
is characterized by two string equations which can be solved  in terms of a \emph{semiclassical} expansion
\begin{equation}\label{eps}
u=\sum_{k\geq 0}\epsilon^k\,u^{(k)}(x,\bt),\quad v=\sum_{k\geq
0}\epsilon^{2k}\,v^{(2k)}(x,\bt).
\end{equation}
The coefficients $(u^{(k)},v^{(2k)})$ can be expressed as rational functions of the leading terms
$(u^{(0)},v^{(0)})$ and their derivatives with respect to $x$.

A method to determine the expansions \eqref{eps} was provided in
\cite{mel1}.
It uses two  functions
\[
\R:=\sum_{k\geq
0}\dfrac{R_k(u,v)}{z^k},\quad
\T:=\sum_{k\geq
0}\dfrac{T_k (u,v)}{z^k},\quad R_0=T_0=1,
\]
which are  related to the resolvent $\mR:=(z-\mL)^{-1}$ of the Lax operator $\mL$ through the equation
\[
\Big(z-\dfrac{2\,\R}{(1+\,\T)}\,\Lambda\Big)\,\mR_+=\R,
\]
so that they generate the trace densities $\mbox{Res}(\mR\,\Lambda^{-j})$ for $j\geq 0$,
where $\mbox{Res}(\sum c_k\,\Lambda^k):=c_0$.  These functions are determined
by the equations
\begin{equation}\label{sis1}
\begin{cases}
\T+\T_{[-1]}+\dfrac{2}{z}(u-z)\,\R=0,\\\\
\T^2-\dfrac{4}{z^2}\,v_{[1]}\,\R\,\R_{[1]}=1.
\end{cases}
\end{equation}
where we are denoting
\[
F_{[r]}(x):=F(x+r\,\epsilon).
\]
As it was shown in
\cite{mel1}  the string equations for $(u,v)$ reduce to the system
\begin{equation}\everymath{\displaystyle}\label{sis2}
\begin{cases}
\oint_{\gamma}\dfrac{d z}{2\pi i\,z}\,
V_z(z,\bt)\,\R(z,u,v)\, =0,\\\\
\oint_{\gamma}\dfrac{d z}{2\pi i}\,V_z(z,\bt)\,\T_{[-1]}(z,u,v)\,
=-2\,x,
\end{cases}
\end{equation} where
$
V(z,\ct):=\sum_{j=1}^\infty
t_j\,z^j,
$
and $\gamma$ is a large positively oriented closed path.
For $\epsilon=0$ the system \eqref{sis2}  reduces to a pair of
hodograph type equations
\begin{equation}\label{hods}
f_v(\bt,u^{(0)},v^{(0)})=0,\quad f_u(\bt,u^{(0)},v^{(0)})=2\,x,
\end{equation}
where
\begin{equation}\label{efe}
f(\bt,u,v):=\frac{1}{2\pi i}\oint_{\gamma}dz\,V_z(z,\bt)\,\left((z-u)^2-4v\right)^{\frac{1}{2}}=-2\,v\,t_1-4\,u\,v\,t_2+\ldots.
\end{equation}
These equations determine $(u^{(0)},v^{(0)})$ and represent the
planar limit contribution to the partition function of the Hermitian
model \cite{gin}.  However, near critical points of \eqref{hods}
the functions $(u^{(0)},v^{(0)})$ are multi-valued and have singular
$x$-derivatives (\emph{gradient catastrophe}),  so that  the
 semiclassical expansion \eqref{eps} breaks down. In terms of the Hermitian matrix model this situation corresponds
 to the critical points of the large $N$ expansion of $Z_N(\bt)$  and, as it is well-known  \cite{ds1}-\cite{mira}, a \emph{double scaling limit
 } method of regularization  is available. This method leads to an important
  nonperturbative approach to two-dimensional quantum gravity.

  Recent research \cite{gra1}-\cite{dub} shows that the double scaling limit expansions are also relevant
in the asymptotic analysis of solutions of dispersionful integrable systems. They supply good approximations in the
 transition regions from the semiclassical to the oscillatory regimes after the time of gradient catastrophe. These
applications motivate the interest in reconsidering the double scaling limit method from the point of view of the
 theory of string equations for integrable systems of  dispersionful type.  The formulation of the double scaling limit within the theory of the dispersionful Toda hierarchy
 was already addressed in \cite{mel1}, but only the case corresponding to $u=0$ and even potentials $V(-z)=V(z)$ was
 analyzed.  The purpose of the present paper is to analyze the general case
that exhibits a much richer structure of critical behaviours (see
\cite{petro1}-\cite{mira} for its relevance in the Hermitian matrix
model of random surfaces and quantum gravity).

 According to the strategy of the double scaling method we
look at the systems \eqref{sis1}-\eqref{sis2} as singular
perturbation problems in the small parameter $\epsilon$ and
 consider the semiclassical expansions near critical points as \emph{outer} expansions of the solutions
 of \eqref{sis1}-\eqref{sis2}. Consequently the problem consists in characterizing appropriate \emph{inner} expansions
\begin{equation}\label{expu1}
x=x_c+\overline{\epsilon}^{2n}\overline{x},\quad u=u_c+\sum_{j\geq2}\overline{\epsilon}^j\overline{u}^{(j)},\quad
v=v_c+\sum_{j\geq2}\overline{\epsilon}^j\overline{v}^{(j)},\quad
\epsilon=\overline{\epsilon}^{2\,n+1}
\end{equation}
with coefficients depending on stretched variables
$(\overline{x},\overline{\bt})$. This type of expansions were determined in
the theory of random matrix models through several methods of
analysis of discrete string equations for orthogonal polynomials in
the large $N$ limit. The present work provides an alternative scheme
based on the characterization of solutions
\[
\R=\sum_{j\geq0}\overline{\epsilon}^j\overline{R}^{(j)},\quad \T=\sum_{j\geq0}\overline{\epsilon}^j\overline{T}^{(j)},
\]
of the resolvent trace equations \eqref{sis1} of the
dispersionful Toda hierarchy. This scheme establishes the existence
of the series \eqref{expu1} to all orders in
$\overline{\epsilon}$ and leads to the characterization of the subleading coefficients $(\overline{u}^{(2)},\overline{v}^{(2)})$ in terms of ordinary differential equations involving the Gel'fand-Dikii polynomials of the
resolvent trace expansion of the Schr\"{o}dinger operator.

We analyze  two different
classes of critical points. Firstly, we consider the sets $\mC_n$
of critical points which underlie the Hermitian matrix model of
quantum gravity \cite{petro1}-\cite{petro2} and we determine inner expansions \eqref{expu1} which satisfy the \emph{doubling property}
$$
c_{\pm}\,G_{n,\pm}(u_{\pm})=\overline{x},\quad u_{\pm}:=v^{(2)}\pm \sqrt{v_c}\,u^{(2)},
$$
with $G_{n,\pm}$ being the Gel'fand-Dikii polynomials. Secondly,
we characterize a wide family of sets $\mC_{lmn}$  of critical points
where  $l\geq 1,\, m=1,\ldots, l$ and  $n\geq 2$,
describe the branching behavior of $(u^{(0)},\,v^{(0)})$ near
the critical points. Then we develop the  double scaling limit
method for the case  $n=2$ with stretched
variables
$$x=x_c+\bar{\epsilon}^4\bar{x},\quad t_j=t_{c,j}+\bar{\epsilon}^4\bar{t}_j,\quad j\geq1, \quad
\bar{\epsilon}=\epsilon^{1/5}.$$
For $\mC_{112}$ we find that the subleading coefficients satisfy
$$\overline{v}^{(2)}=\pm\,\sqrt{v_c}\,\overline{u}^{(2)},$$
with $\overline{v}^{(2)}$ being
determined by the Painlev\'e I equation. For the other cases $\mC_{lm2},\,
l\geq 2$, we find again the doubling property. Indeed, the functions
\[
u_{\pm}:=\overline{v}^{(2)}\pm\,\sqrt{v_c}\,\overline{u}^{(2)},
\]
turn out to be  determined by a pair of decoupled Painlev\'e I equations.

The  double scaling limit expansions at critical points of
the types
$\mC_{lmn}$ can be applied to the asymptotic analysis of solutions of the dispersionful Toda hierarchy.
 As an illustrative example we end our work by applying our results to regularize some critical processes of ideal  models of
 Hele-Shaw flows \cite{lee3}-\cite{lee2} based on the Toda hierarchy.

Our scheme can be  formulated with other scaling ans\"{a}tze for stretched variables like those of Galilean type used in
\cite{gra1}-\cite{dub}. Furthermore, it can be applied to the study of the double
scaling limit
of the normal matrix model which is related to another solution of the two-dimensional dispersionful Toda hierarchy \cite{zab1}-\cite{zab3} determined by
a pair of  string equations. We also expect the methods of this paper to be applicable to dispersionful versions of the multi-component KP hierarchies which arise in random matrix theory and Dyson Brownian motions \cite{avm}.

The layout of the paper is as follows: in section 2 we show how inner expansions of the generating functions $\R$ and $\T$
can be determined from the resolvent trace equations \eqref{sis1}. In section 3 we  characterize
the double scaling limit expansions at critical points satisfying strong conditions. Finally, we devote section 4 to introduce
families $\mC_{lmn}$  of critical points and to develop the corresponding double scaling limit method.

\section{Inner expansions of the resolvent functions}

Let us consider the hodograph system \eqref{hods}
\begin{equation}\label{hods2}
f_v=0,\quad f_u=2\,x.
\end{equation}
Notice that the function $f$ defined in \eqref{efe} satisfies
\begin{equation}\label{pde}
f_{uu}-v\,f_{vv}=0.
\end{equation}
The hodograph system determines a solution of the dispersionless Toda
hierarchy provided that  the implicit function theorem applies for obtaining $(u(x,\bt),v(x,\bt))$ from \eqref{hods2}.
 For example \eqref{hods2} implies
\[
u_{t_1}=v_x,\quad v_{t_1}=v\,u_x,
\]
so that
\[
v_{xx}-(\log v)_{t_1 t_1}=0,
\]
which is the hyperbolic version of the long wave limit of the Toda lattice equation. It is also easy to
see that the solutions
of \eqref{hods2} verify
\[
u_{t_2}=2\,(u\,u_{t_1}+\,v_{t_1}),\quad v_{t_2}=2\,(v\,u)_{t_1},
\]
which is the dispersionless limit of the focusing nonlinear Schr\"{o}dinger equation \cite{dub}.

The set of critical points $(x_c,\bt_c,u_c,v_c)$ of \eqref{hods2} is characterized by the system
\begin{equation}\label{1}
f_v=0,\quad f_u=2\,x\quad
f_{uv}^2-f_{uu}f_{vv}=0,
\end{equation}
Our main aim in this work is to characterize  inner
expansions for solutions $(u,v)$ of the system \eqref{sis1}-\eqref{sis2} near  critical points
of \eqref{hods2}.

\subsection{Symmetric variables}

Let us start by
analyzing the resolvent trace system \eqref{sis1}. It is form invariant under the change
of variables
\begin{equation}\label{nev}
x=x_c+\overline{\epsilon}^{2n}\overline{x},\quad
\epsilon=\overline{\epsilon}^{2\,n+1},\quad n\geq 2.
\end{equation}
Indeed since $\epsilon\,\partial _x=\overline{\epsilon}\,\partial
_{\overline{x}}$ we  can  rewrite  \eqref{sis1} as
\begin{equation}\everymath{\displaystyle}\label{sis1b}
\begin{cases}
\T+\T_{[\overline{-1}]}+\dfrac{2}{z}(u-z)\,\R=0,\\\\
\T^2-\dfrac{4}{z^2}\,v_{[\overline{1}]}\,\R\,\R_{[\overline{1}]}=1,
\end{cases}
\end{equation}
where we are denoting
\[
F_{[\bar{r}]}(\bar{x}):=F(\bar{x}+r\,\bar{\epsilon}).
\]
We will prove that if $(u,v)$ are expansions of the form
\begin{equation}\label{expu}
u=u_c+\sum_{j\geq2}\overline{\epsilon}^j\overline{u}^{(j)},\quad
v=v_c+\sum_{j\geq2}\overline{\epsilon}^j\overline{v}^{(j)},
\end{equation}
then the solution of \eqref{sis1b} can be expressed as
\begin{equation}\label{critexp}\everymath{\displaystyle}
\T=\sum_{j\geq0}\overline{\epsilon}^j\overline{T}^{(j)},\quad
\R=\sum_{j\geq0}\overline{\epsilon}^j\overline{R}^{(j)},
\end{equation}
where the coefficients $(\overline{T}^{(j)},\,\overline{R}^{(j)})$ are differential polynomials in the coefficients of the expansions of $(u,v)$.

An important property of the combined system  \eqref{sis1}-\eqref{sis2} is its invariance under the transformation
\begin{align}\label{sym}
\nonumber &\widehat{u}(\overline{\epsilon},\overline{x}):=u(-\overline{\epsilon},\overline{x}+\overline{\epsilon}),\quad \widehat{v}(\overline{\epsilon},\overline{x}):=v(-\overline{\epsilon},\overline{x}),\\  \\
\nonumber  &
\widehat{\T}(\overline{\epsilon},z,\overline{x}):=\T(-\overline{\epsilon},z,\overline{x}+2\,\overline{\epsilon}),\quad
\widehat{\R}(\overline{\epsilon},z,\overline{x}):=\R(-\overline{\epsilon},z,\overline{x}+\overline{\epsilon}).
\end{align}
Thus it is convenient to introduce the \emph{symmetric} variables
\begin{equation}\label{siv}
w(\overline{\epsilon}):=u(\overline{\epsilon},\overline{x}-\frac{\overline{\epsilon}}{2})=u_{[\overline{-1/2}]}(\overline{\epsilon}),
\end{equation}
\begin{equation}\label{siv1}
\U(\overline{\epsilon}):=\T(\overline{\epsilon},z,\overline{x}-\overline{\epsilon})=\T_{[\overline{-1}]}(\overline{\epsilon})
\quad \V(\overline{\epsilon},z,\overline{x}):=\R(\overline{\epsilon},z,\overline{x}-\frac{\overline{\epsilon}}{2})=R_{[\overline{-1/2}]}(\overline{\epsilon}),
\end{equation}
which transform under \eqref{sym} in the simple form
\begin{equation}\label{simw}
\widehat{w}(\overline{\epsilon})=w(-\overline{\epsilon}),\quad
\widehat{\U}(\overline{\epsilon}):=\U(-\overline{\epsilon}),\quad
\widehat{\V}(\overline{\epsilon})=\V(-\overline{\epsilon}).
\end{equation}
In terms of $(w,v,\U,\V)$ the system \eqref{sis1b} reads

\begin{equation}\everymath{\displaystyle}\label{sis1c}
\begin{cases}
\U_{[\overline{1}]}+
\U+\frac{2}{z}\,(w_{[\overline{1/2}]}-z)\,
\V_{[\overline{1/2}]}=0,
\\ \\
\U^2-\frac{4}{z^2}\,v\,
\V_{[\overline{-1/2}]}\,
\V_{[\overline{1/2}]}=1.
\end{cases}
\end{equation}
Moreover, from \eqref{sis1c} one easily deduces the following linear equation for $\U$
\begin{align}\label{lin}
\nonumber (z-w_{[\overline{-1/2}]})&\,(z-w_{[\overline{1/2}]})\,(z-w_{[\overline{3/2}]})\,(\U_{[\overline{1}]}-\U)=
v_{[\overline{1}]}\,(z-w_{[\overline{-1/2}]})\,(\U_{[\overline{2}]}+
\U_{[\overline{1}]})
\\\\
\nonumber &-v\,(z-w_{[\overline{3/2}]}))\,(\U_{[\overline{-1}]}+
\U).
\end{align}
It can be expressed as
\begin{equation}\label{lin1}
\Big(\sum_{k\geq 1}\overline{\epsilon}^k\, \mathcal{J}_k\Big)\,\U=0,
\end{equation}
where $\mathcal{J}_k$ are linear differential operators depending on the coefficients of the expansions of $(w,v)$ of the form
\[
\mathcal{J}_k=(z-u_c)\,r_c\,\mathcal{J}_{k,0}+(z-u_c)\,\mathcal{J}_{k,1}+r_c\,\mathcal{J}_{k,2}+\mathcal{J}_{k,3},
\]
where
\[
r_c=r_c(z):=(z-u_c)^2-4\,v_c,
\]
and  $\mathcal{J}_{k,i}$ are $z$-independent. For example, the first few are
\begin{align}\label{jot}
\nonumber
& \mathcal{J}_1=(z-u_c)\,r_c\,\partial_{\overline{x}},\quad \mathcal{J}_2=\dfrac{1}{2}\,(z-u_c)\,r_c\,\partial_{\overline{x}}^2,\\\\
\nonumber
&\mathcal{J}_3=\dfrac{z-u_c}{6}\,r_c\,\partial_{\overline{x}}^3-3\,r_c\,u^{(2)}\,\partial_{\overline{x}}-
(z-u_c)\,(v_c\,\partial_{\overline{x}}^3+4\,v^{(2)}\,\partial_{\overline{x}}+
2\,v^{(2)}_{\overline{x}})-4\,v_c\,(2\,u^{(2)}\,\partial_{\overline{x}}+\,u^{(2)}_{\overline{x}}).
\end{align}
We will now use both \eqref{sis1c} and \eqref{lin} to determine the expansions of $\U$ and $\V$.

If we substitute in \eqref{sis1c} the expansions
\begin{align*}
&w=u_c+\sum_{j\geq2}\overline{\epsilon}^j\overline{w}^{(j)},\quad
v=v_c+\sum_{j\geq2}\overline{\epsilon}^j\overline{v}^{(j)},
\\
&
\V=\sum_{j\geq0}\overline{\epsilon}^j\overline{V}^{(j)},\quad
\U=\sum_{j\geq0}\overline{\epsilon}^j\overline{U}^{(j)},
\end{align*}
then by identifying the coefficients of the powers of $\overline{\epsilon}$ we find
$$\frac{\overline{V}^{(0)}}{z}=\frac{1}{r_c^{1/2}},\quad
\overline{U}^{(0)}=\frac{z-u_c}{r_c^{1/2}},\quad
\frac{\overline{V}^{(1)}}{z}=\overline{U}^{(1)}=0,$$
and the recurrence system $(j\geq2)$
$$\everymath{\displaystyle}\begin{array}{lll}
\overline{U}^{(j)}-(z-u_c)\frac{\overline{V}^{(j)}}{z}&=&
-\frac{1}{2}\sum_{k+i=j,\,k\geq1}\frac{1}{k!}\partial_{\overline{x}}^{k}\overline{U}^{(i)}+
(z-u_c)\sum_{k+i=j,\,k\geq1}\frac{1}{2^{k}(k)!}
\partial_{\overline{x}}^{k}\frac{\overline{V}^{(i)}}{z}-\\  \\
  &   &\sum_{\scriptsize\begin{array}{c} i_1+i_2+k_1+k_2=j\\ i_1\geq2\end{array}}
\frac{1}{2^{k_1+k_2}k_1!k_2!}(\partial_{\overline{x}}^{k_1}\overline{w}^{(i_1)})
\left(\partial_{\overline{x}}^{k_2}\frac{\overline{V}^{(i_2)}}{z}\right),\\  \\
\overline{U}^{(0)}\overline{U}^{(j)}-4v_c\frac{\overline{V}^{(0)}}{z}\frac{\overline{V}^{(j)}}{z}&=&
-\frac{1}{2}\sum_{i+k=j,\,1\leq i,k\leq j-1}\overline{U}^{(i)}\overline{U}^{(k)}+\\  \\
  &  &\sum_{\scriptsize\begin{array}{c} i_1+i_2+k_1+k_2=j\\ 1\leq i_1,i_2\leq j-1,\,k_1+k_2\,\mbox{even}\end{array}}
\frac{(-1)^{k_1}}{2^{k_1+k_2-1}k_1!k_2!}v_c\left(\partial_{\overline{x}}^{k_1}\frac{\overline{V}^{(i_1)}}{z}\right)
\left(\partial_{\overline{x}}^{k_2}\frac{\overline{V}^{(i_2)}}{z}\right)+\\  \\
  &  &\sum_{\scriptsize\begin{array}{c} i_1+i_2+k_1+k_2+l=j\\ l\geq2,\,k_1+k_2\,\mbox{even}\end{array}}
\frac{(-1)^{k_1}}{2^{k_1+k_2}k_1!k_2!}\overline{v}^{(l)}\left(\partial_{\overline{x}}^{k_1}\frac{\overline{V}^{(i_1)}}{z}\right)
\left(\partial_{\overline{x}}^{k_2}\frac{\overline{V}^{(i_2)}}{z}\right)
\end{array}$$
These equations are linear with respect to   $\overline{U}^{(j)}$ and $\dfrac{\overline{V}^{(j)}}{z}$. Furthermore the determinant of their corresponding coefficients is $r_c^{1/2}$.
\newpage
Thus one obtains
$$\everymath{\displaystyle}\begin{array}{l}
\overline{U}^{(2)}=\frac{1}{r_c^{\frac{3}{2}}}[4v_c\overline{w}^{(2)}+2(z-u_c)\overline{v}^{(2)}],\quad
\frac{\overline{V}^{(2)}}{z}=\frac{1}{r_c^{\frac{3}{2}}}[2\overline{v}^{(2)}+(z-u_c)\overline{w}^{(2)}],\\  \\
\overline{U}^{(3)}=\frac{1}{r_c^{\frac{3}{2}}}[4v_c\overline{w}^{(3)}+2(z-u_c)\overline{v}^{(3)}],\quad
\frac{\overline{V}^{(3)}}{z}=\frac{1}{r_c^{\frac{3}{2}}}[2\overline{v}^{(3)}+(z-u_c)\overline{w}^{(3)}],\\  \\
\overline{U}^{(4)}=\frac{1}{r_c^{\frac{3}{2}}}\left[4v_c\overline{w}^{(4)}+\frac{v_c}{2}\overline{w}^{(2)}_{\overline{x}\overline{x}}+
4\overline{v}^{(2)}\overline{w}^{(2)}+2(z-u_c)\overline{v}^{(4)}\right]+\\  \\
\qquad\qquad+\frac{1}{r_c^{\frac{5}{2}}}\left[4v_c^2\overline{w}^{(2)}_{\overline{x}\overline{x}}
+24v_c\overline{v}^{(2)}\overline{w}^{(2)}+
(z-u_c)\left(2v_c\overline{v}^{(2)}_{\overline{x}\overline{x}}+6\overline{v}^{(2)^2}+6v_c\overline{w}^{(2)^2}\right)\right],\\   \\
\frac{\overline{V}^{(4)}}{z}=\frac{1}{r_c^{\frac{3}{2}}}\left[2\overline{v}^{(4)}+
\frac{1}{4}\overline{v}^{(2)}_{\overline{x}\overline{x}}+\overline{w}^{(2)^2}+(z-u_c)\overline{w}^{(4)}\right]+\\  \\
\qquad\qquad\frac{1}{r_c^{\frac{5}{2}}}\left[2v_c\overline{v}^{(2)}_{\overline{x}\overline{x}}+6v_c\overline{w}^{(2)^2}+
6\overline{v}^{(2)^2}+(z-u_c)\left(v_c\overline{w}^{(2)}_{\overline{x}\overline{x}}+6\overline{v}^{(2)}\overline{w}^{(2)}\right)\right],
\end{array}$$
and by using induction one proves that the coefficients of the expansions of $\U$ and $\V$ are of the form
\begin{align}\label{ues}
\nonumber &\overline{U}^{(2j)}=
 \sum_{l=1}^j\frac{\alpha_{l}^{(2j)}(z-u_c)+\beta_{l}^{(2j)}}{r_c^{l+\frac{1}{2}}},\quad j\geq3;\quad
\overline{U}^{(2j+1)}=\sum_{l=1}^j\frac{\alpha_{l}^{(2j+1)}(z-u_c)+\beta_{l}^{(2j+1)}}{r_c^{l+\frac{1}{2}}},\quad j\geq2,\\  \\
\nonumber &\dfrac{\overline{V}^{(2j)}}{z}=\sum_{l=1}^j\frac{\gamma_{2j\,l}(z-u_c)+\eta_{2j\,l}}{r_c^{l+\frac{1}{2}}},\quad j\geq3;\quad  \dfrac{\overline{V}^{(2j+1)}}{z}=
\sum_{l=1}^j\frac{\gamma_{l}^{(2j+1)}(z-u_c)+\eta_{l}^{(2j+1)}}{r_c^{l+\frac{1}{2}}},\quad j\geq2.
\end{align}
Moreover by substituting \eqref{ues} in the linear equation \eqref{lin} and taking into account \eqref{jot} it can be proved by induction that the functions  $\alpha_{l}^{(i)}$, $\beta_{l}^{(i)}$, $\gamma_{l}^{(i)}$ and $\eta_{l}^{(i)}$ are differential polynomials in
$(\overline{v}^{(2)}$,...,$\overline{v}^{(i-2\,l+2)}, \overline{w}^{(2)}$,...,$\overline{w}^{(i-2\,l+2)})$.

\subsection{Invariant solutions}

The solution $(\U,\V)$ of the system \eqref{sis1c} is uniquely determined by  $(w,v)$. Hence if we assume that $(w,v)$ are even functions
of $\overline{\epsilon}$, then as a consequence of the invariance of \eqref{sis1c} under the transformation \eqref{sym} we deduce
that $(\U,\V)$ are even functions of $\overline{\epsilon}$ too. In this way we have
\begin{equation}\label{inva}
w=u_c+\sum_{j\geq1}\overline{\epsilon}^{2\,j}\overline{w}^{(2\,j)},\quad
v=v_c+\sum_{j\geq1}\overline{\epsilon}^{2\,j}\overline{v}^{(2\,j)},
\quad
\U=\sum_{j\geq0}\overline{\epsilon}^{2\,j}\overline{U}^{(2\,j)},\quad
\V=\sum_{j\geq0}\overline{\epsilon}^{2\,j}\overline{V}^{(2\,j)}.
\end{equation}
These solutions will be henceforth called \emph{invariant solutions} of the resolvent trace equations.

As we have seen in the above subsection the coefficients $\overline{U}^{(2j)}$ and $\overline{V}^{(2j)}$ can be expanded in powers of $r_c$ with
leading terms
\begin{equation}\label{leadi}
\overline{U}^{(2j)}=\frac{(z-u_c)\,\alpha_{j}^{(2j)}+\beta_{j}^{(2j)}}{r_c^{j+\frac{1}{2}}}+\mathcal{O}
\Big(\frac{1}{r_c^{j-\frac{1}{2}}}\Big), \quad
\frac{\overline{V}^{(2j)}}{z}=\frac{(z-u_c)\,\gamma_{j}^{(2j)}+\eta_{j}^{(2j)}}{r_c^{j+\frac{1}{2}}}+\mathcal{O}
\Big(\frac{1}{r_c^{j-\frac{1}{2}}}\Big).
\end{equation}
Furthermore, by identifying the coefficient of $\overline{\epsilon}^{2j}$ in the first equation of the system
\eqref{sis1c} and by taking into account that $(z-u_c)^2=r_c+4\,v_c$ it follows that
\begin{equation}\label{leadi1}
\gamma_{j}^{(2j)}=\dfrac{\beta_{j}^{(2j)}}{4\,v_c},\quad \eta_{j}^{(2j)}=\alpha_{j}^{(2j)}.
\end{equation}
Now we prove the connection between
 these coefficients
and the Gel'fand-Dikii differential polynomials of the KdV theory.
\begin{teh} The functions
\begin{equation}\label{gdp}
G_{j,\pm}:=\dfrac{\beta_{j}^{(2j)}}{2\,v_c}\pm\dfrac{\alpha_{j}^{(2j)}}{\sqrt{v_c}},
\end{equation}
are the Gel'fand-Dikii differential polynomials in $u_{\pm}:=v^{(2)}\pm \sqrt{v_c}\,u^{(2)}$, respectively.
\end{teh}
\begin{proof}
By identifying the coefficient of $\overline{\epsilon}^{2j+1}$ in \eqref{lin1} we get
\begin{equation}\label{jot1}
\sum_{k+2l=2j+1}\mathcal{J}_k\,\overline{U}^{(2l)}=0.
\end{equation}
From \eqref{jot} and by taking into account that $(z-u_c)^2=r_c+4\,v_c$ it is clear that only the terms  $\mathcal{J}_1\overline{U}^{(2j)}$ and $\mathcal{J}_3\overline{U}^{(2j-2)}$ contribute to the coefficient of
$\frac{1}{r_c^{j-\frac{1}{2}}}$ in \eqref{jot1}. Thus
we get the recursion relations
\begin{align}\label{rec}
\nonumber &
\partial_{\overline{x}}\,\alpha_{j}^{(2j)}=(v_c\,\partial_{\overline{x}}^3+4\,\overline{v}^{(2)}\,
\partial_{\overline{x}}+2\,\overline{v}^{(2)}
_{\overline{x}})\,\alpha_{j-1}^{(2(j-1))}+(2\,\overline{u}^{(2)}\,\partial_{\overline{x}}+\overline{u}^{(2)}_{\overline{x}})\,
\beta_{j-1}^{(2(j-1))},\\\\
\nonumber &
\partial_{\overline{x}}\,\beta_{j}^{(2j)}=(v_c\,\partial_{\overline{x}}^3+4\,\overline{v}^{(2)}\,\partial_{\overline{x}}
+2\,\overline{v}^{(2)}
_{\overline{x}})\,\beta_{j-1}^{(2(j-1))}+4\,v_c\,(2\,\overline{u}^{(2)}\,\partial_{\overline{x}}+\overline{u}^{(2)}_{\overline{x}})\,
\alpha_{j-1}^{(2(j-1))},
\end{align}
which lead at once to the well-known third-order differential equation for the Gel´fand-Dikii differential polynomials
\begin{equation}\label{thir}
\partial_{\overline{x}}\,G_{j,\pm}=
(v_c\,\partial_{\overline{x}}^3+4\,u_{\pm}\,\partial_{\overline{x}}+2\,
u_{\pm,\overline{x}})\,\,G_{j-1,\pm}.
\end{equation}
\end{proof}

From \eqref{gdp} we have that $G_{1,\pm}=\pm\frac{2}{\sqrt{v_c}}u_{\pm}$ and by using \eqref{thir}
we find
$$\everymath{\displaystyle}\begin{array}{lll}
G_{2,\pm}&=&\pm\frac{2}{\sqrt{v_c}}\left(v_cu_{\pm,\overline{x}\overline{x}}+
3u_{\pm}^2\right),\\  \\
G_{3,\pm}&=&\pm\frac{2}{\sqrt{v_c}}\left(v_c^2u_{\pm,\overline{x}\overline{x}\overline{x}\overline{x}}
+10v_cu_{\pm,\overline{x}\overline{x}}u_{\pm}+5v_cu_{\pm,\overline{x}}^2+
10u_{\pm}^3\right),\\  \\
G_{4,\pm}&=&\pm\frac{2}{\sqrt{v_c}}\Big(v_c^3u_{\pm,\overline{x}\overline{x}\overline{x}\overline{x}\overline{x}
\overline{x}}+
14v_c^2u_{\pm}u_{\pm,\overline{x}\overline{x}\overline{x}\overline{x}}+28v_c^2u_{\pm,\overline{x}\overline{x}\overline{x}}
u_{\pm,\overline{x}}+
21v_c^2u_{\pm,\overline{x}\overline{x}}^2\\  \\
  &  &\quad+70v_cu_{\pm}^2u_{\pm,\overline{x}\overline{x}}+70v_cu_{\pm}u_{\pm,\overline{x}}^2+35u_{\pm}^4\Big).
\end{array}$$

\section{Strong conditions for critical points and the doubling property}

In what follows the notation $F^{(c)}$ will represent the value  of a function $F$ at a critical point
$(x_c,\bt_c,u_c,v_c)$ of \eqref{hods2}. We will also suppose that $v_c\neq 0$.
The following sets of critical points  were considered in the applications of the Hermitian matrix model to quantum gravity \cite{mira}-\cite{petro1}.

\begin{defi} Given $n\geq 2$ we denote by $\mC_{n}$ the set of critical points of the hodograph system \eqref{hods2}
which satisfy the (\emph{strong}) conditions
\begin{equation}\label{strong}
(\partial^k_v\,f)^{(c)}=(\partial_u\,\partial^{l}_v\,f)^{(c)}=0,\quad
1\leq k\leq n,\quad 1\leq l\leq n-1,
\end{equation}
and such that $((\partial^{n+1}_v\,f)^{(c)},(\partial_u\,\partial^{n}_v\,f)^{(c)})
\neq (0,0)$.
\end{defi}

Due to \eqref{pde} it is clear that $\mC_{n}$ is also determined by the condition that all derivatives $(\partial^k_u\,\partial^l_v\,f)^{(c)}$ with $(k,l)\neq (1,0)$ up to order $n$ vanish and such that at least one $n+1$-order derivative is different from zero.

An alternative characterization of $\mC_{n}$ can  be formulated in terms of the integrals
\begin{equation}\label{iji}
I_k(\bt_c,u_c,v_c):=\frac{1}{2\pi i}\oint_{\gamma}dz\,\dfrac{V_z(z,\bt_c)}{r_c^{k\,+1/2}},\quad
J_k(\bt_c,u_c,v_c):=\frac{1}{2\pi i}\oint_{\gamma}dz\,(z-u_c)\,\dfrac{V_z(z,\bt_c)}{r_c^{k\,+1/2}}.
\end{equation}
Indeed, $I_k$ and $J_k$ are proportional to $(\partial^{k+1}_v\,f)^{(c)}$ and $(\partial_u\,\partial^{k}_v\,f)^{(c)}$ respectively. Hence it follows that

\begin{Lemma} $(x_c,\bt_c,u_c,v_c)\in \mC_{n}$ if and only if
\begin{equation}\label{criti}
I_0=0,\,\,J_0=-2\,x_c,\quad
I_k=J_k=0,\quad \mbox{for $1\leq k\leq n-1$ },\quad (I_n,J_n)\neq (0,0).
\end{equation}
\end{Lemma}

 Let us consider the system of  string equations \eqref{sis2} at points $(x,\bt_c,u,v)$ near a given critical point
 $(x_c,\bt_c,u_c,v_c)\in \mC_n$. In terms of symmetric variables it reads
\begin{equation}\everymath{\displaystyle}\label{sis2a}
\begin{cases}
\oint_{\gamma}\dfrac{d z}{2\pi i\,z}\,
V_z(z,\bt_c)\,\V(z)\, =0,\\\\
\oint_{\gamma}\dfrac{d z}{2\pi i}\,V_z(z,\bt_c)\,\U(z)\,
=-2\,x.
\end{cases}
\end{equation}
Then if we set
\[
x=x_c+\overline{\epsilon}^{2n}\,\overline{x},\quad
\epsilon=\overline{\epsilon}^{2\,n+1},
\]
and assume \eqref{inva} we obtain a recursive method for determining the coefficients of $w$ and $v$. Indeed \eqref{sis2a} is equivalent to
\begin{equation}\everymath{\displaystyle}\label{sis2bb}
\begin{cases}
\oint_{\gamma}\dfrac{d z}{2\pi i\,z}\,
V_z(z,\bt_c)\,\overline{V}^{(2j)}(z)\, =0,\\\\
\oint_{\gamma}\dfrac{d z}{2\pi i}\,V_z(z,\bt_c)\,\overline{U}^{(2j)}(z)\,
=-2\,x_c\,\delta_{j0}-2\,\delta_{j\,n}\,\overline{x}.
\end{cases}
\end{equation}
For $0\leq j \leq n-1$ these equations are identically satisfied because of \eqref{criti}. For $j=n$ we get from \eqref{leadi}-\eqref{leadi1}  that
the equations \eqref{sis2bb} reduce to
\begin{equation}\everymath{\displaystyle}\label{sis2c}
\begin{cases}
J_n\,\beta_n^{(2n)}+4\,v_c\,I_n\,\alpha_n^{(2n)} =0,\\\\
I_n\,\beta_n^{(2n)}+J_n\,\alpha_n^{(2n)}=-2\,\overline{x},
\end{cases}
\end{equation}
or, equivalently, in terms of the Gel'fand-Dikii polynomials \eqref{gdp} we obtain a pair of decoupled ordinary differential equations for
$u_{\pm}:=v^{(2)}\pm \sqrt{v_c}\,u^{(2)}$
\begin{equation}\label{gdpe}
\Big(J_n\pm2\,\sqrt{v_c}\,I_n\Big)\,G_{n,\pm}(u_{\pm})=\mp\,\frac{2}{\sqrt{v_c}}\,\overline{x},
\end{equation}
Thus if the condition
\[
J_n\pm2\,\sqrt{v_c}\,I_n\neq 0,
\]
is satisfied, we have the so-called \emph{doubling property} arising in the one-cut case of the Hermitian matrix
model  \cite{petro1}-\cite{petro2}.

 The first few cases of \eqref{gdpe} are
$$\begin{array}{l}
(f_{uvv}^{(c)}\pm\sqrt{v_c}\,f_{vvv}^{(c)})\left(v_c\,u_{\pm,\overline{x}\overline{x}}+
3\,u_{\pm}^2\right)=12\,\overline{x},\quad n=2\,;\\  \\
(f_{uvvv}^{(c)}\pm\sqrt{v_c}\,f_{vvvv}^{(c)})\left(v_c^2\,
u_{\pm,\overline{x}\overline{x}\overline{x}\overline{x}}
+10\,v_c\,u_{\pm,\overline{x}\overline{x}}u_{\pm}+5\,v_c\,u_{\pm,\overline{x}}^2+
10\,u_{\pm}^3\right)=120\,\overline{x},\quad n=3\,;\\  \\
(f_{uvvvv}^{(c)}\pm\sqrt{v_c}\,f_{vvvvv}^{(c)})\Big(v_c^3\,u_{\pm,\overline{x}\overline{x}\overline{x}\overline{x}\overline{x}\overline{x}}+
14\,v_c^2\,u_{\pm}\,u_{\pm,\overline{x}\overline{x}\overline{x}\overline{x}}+28\,v_c^2\,u_{\pm,\overline{x}\overline{x}\overline{x}}u_{\pm,\overline{x}}+
21\,v_c^2\,u_{\pm,\overline{x}\overline{x}}^2\\  \\
\qquad\qquad+70\,v_c\,u_{\pm}^2\,u_{\pm,\overline{x}\overline{x}}+
70\,v_c\,u_{\pm}\,u_{\pm,\overline{x}}^2+35\,u_{\pm}^4\Big)=1680\,\overline{x},\quad n=4.
\end{array}$$

Finally for $j>n$ the system \eqref{sis2bb} yields the pair of equations
\begin{equation}\everymath{\displaystyle}\label{sis2d}
\sum_{l=n}^j\,\Big(J_l\,\gamma_l^{(2j)}+I_l\,\eta_l^{(2j)}\Big) =0,\quad
\sum_{l=n}^j\,\Big(J_l\,\alpha_l^{(2j)}+I_l\,\beta_l^{(2j)}\Big)=0,
\end{equation}
which determine each pair $(w^{(2(j-n+1)},v^{(2(j-n+1)})$ recursively.

\section{Further critical points and regularized expansions}

Let us go back to the system \eqref{1} for critical points of the hodograph equations \eqref{hods2}. It  is equivalent to
\begin{equation}\label{3c}
f_v=0,\quad f_u=2\,x,\quad
f_{uv}=\sigma\,\sqrt{v}\,f_{vv},\quad \sigma=\pm 1.
\end{equation}
Let us consider solutions of the hodograph equations near critical points and
assume that
\[
f_v(\bt_c,u,v_c)\not\equiv 0
\]
(similar results are obtained by interchanging the roles of $u$ and $v$). Then there exists an integer $l\geq 1$  verifying
\begin{equation}\label{w2}
(\partial^k_u\,f_v)^{(c)}=0,\, (0\leq k<l),\quad
(\partial^l_u\,f_v)^{(c)}\neq 0.
\end{equation}
As a consequence the first
hodograph equation
$f_v(\bt_c,u,v)=0$ can be used to eliminate the variable $u$  as a function of $v$ near $(u_c,v_c)$. Indeed, from
Weierstrass' preparation theorem it follows that near $(u_c,v_c)$ there exists a factorization
\begin{equation}\label{w3}
f_v(\bt_c,u,v)=(A_0(\bt_c,v)+\ldots+A_{l-1}(\bt_c,v)\,u^{l-1}+u^l)\,
g(\bt_c,u,v),
\end{equation}
where $A_j\, (0\leq j<l)$ are analytic functions of $v$,  and $g$ is
an analytic function of $(u,v)$ which does not vanish near
$(u_c,v_c)$. Hence the solutions of the hodograph equation
$f_v(\bt_c,u,v)=0$ near
$(u_c,v_c)$ are  given by the roots of the polynomial
factor in \eqref{w3}
\begin{equation}\label{w2a}
A_0(\bt_c,v)+\ldots+A_{l-1}(\bt_c,v)\,u^{l-1}+u^l=0,
\end{equation}
and consequently they are characterized by a Puiseux series
\begin{equation}\label{pui}
u(\bt_c,v)=u_c+\sum_{k\geq 1}\,a_k(\bt_c)\,(v-v_c)^{\frac{k}{m}},
\end{equation}
for a certain integer $1\leq m \leq l$.
If we now introduce the function
\begin{equation}\label{w4}
H(\bt_c,w):=f_u(\bt_c,u(\bt_c,v_c+w^m), v_c+w^m),\quad (w:=(v-v_c)^{\frac{1}{m}})
\end{equation}
then at $\bt=\bt_c$ the second hodograph equation in \eqref{hods2}
reads $H(\bt_c,w)=2\,x$.

Furthermore it is easy to see  that as a
consequence of the system \eqref{3c} we have
\[
H_w^{(c)}=f_{uu}^{(c)}\,u_w^{(c)}+f_{uv}^{(c)}\,v_w^{(c)}=\sigma\,\sqrt{v_c}\,f_{vv}^{(c)}\,(\sigma\,\sqrt{v_c}\,u_w^{(c)}+v_w^{(c)}).
\]
On the other hand by differentiating the identity $f_v(\bt_c,u(\bt_c,v_c+w^m), v_c+w^m)\equiv 0$ we get
\[
f_{vu}^{(c)}\,u_w^{(c)}+f_{vv}^{(c)}\,v_w^{(c)}=f_{vv}^{(c)}\,(\sigma\,\sqrt{v_c}\,u_w^{(c)}+v_w^{(c)})=0.
\]
Hence, we deduce that
\begin{equation}\label{hwc}
H_w^{(c)}=0.
\end{equation}
In this way if we assume that
 there exists an integer $n\geq 2$ such that
\begin{equation}\label{w5}
(\partial^k_w\,H)^{(c)}=0,\, (1\leq k<n),\quad
(\partial^n_w\,H)^{(c)}\neq 0,
\end{equation}
then the hodograph equation
$H(\bt_c,w)=2\,x$ determines $w$ as  a function of $x$ with a branch point of order $n-1$ at $x=x_c$

\begin{defi}
We will denote by $\mC_{lmn},\,(l\geq 1,\, 1\leq m \leq l,\, n\geq 2)$ the set of critical points of $\eqref{hods2}$
characterized by \eqref{3c}-\eqref{w2} and such that
\begin{enumerate}
\item $u=u(\bt_c,v)$ has branching order $m-1$  at  $v=v_c$.
\item The corresponding function $H(\bt_c,w)$ defined by \eqref{w4} satisfies \eqref{w5}.
\end{enumerate}
\end{defi}
\subsubsection*{Example }
If we set  $t_n=0$ for $n\neq1,3$ then the system \eqref{3c}  reads
\begin{equation}\label{crite}
t_{c1}+3t_{c3}\,(u_c^2+2\,v_c)=0,\quad 6\,t_{c3}\,u_c\,v_c+x_c=0,\quad u_c=\sigma\, \sqrt{v_c},\quad \sigma=\pm 1.
\end{equation}
Let us consider the critical
points
$(x_c,t_{c1},t_{c3},u_c,v_c)$ with $t_{c3}\neq 0$. They are given by
\[
v_c=-\dfrac{t_{c1}}{9\, t_{c3}},\quad u_c=\sigma\, \sqrt{v_c},\quad ,
\]
where $(x_c,t_{c1},t_{c3})$ are constrained by the equation
\[
3\,x_c=2\,\sigma \, t_{c1}\,\Big(-\dfrac{t_{c1}}{9\, t_{c3}}\Big)^{1/2}.
\]
As we are  assuming that $v_c\neq 0$, we have that
 $t_{c1}\neq 0$ and $u_c\neq 0$ so that \eqref{w2}  is satisfied by $l=1$. Moreover the hodograph equation
\begin{equation}\label{bb}
t_{c1}+3t_{c3}\,(u^2+2\,v)=0,
\end{equation}
leads to
\[
u(\bt_c,w)=\Big(-2\Big(w-\dfrac{v_c}{2}\Big)\Big)^{1/2},\quad w:=v-v_c,\quad v_c=-\dfrac{t_{c1}}{9\, t_{c3}}\neq 0.
\]
Furthermore  $H(\bt_c,w):=-12\,t_{c3}\,u(\bt_c,w)\,(w+v_c)$ verifies $H_{ww}(\bt_c,0)\neq 0$. Therefore, it follows that
$
(x_c,t_{c1},t_{c3},u_c,v_c)\in\mC_{112}.
$

\vspace{0.4cm}

The following statements will be useful for the subsequent discussion.  They are easily proved by
differentiating \eqref{w4} and   the identity
 $f_v(\bt,u(w),v_c+w^{m})=0$.

\begin{Lemma} Given $(x_c,\bt_c,u_c,v_c)\in\mC_{lmn}$ with $v_c\neq0$ then

\begin{itemize}

\item If $l=1$ then  $f_{vv}^{(c)}$ does not vanish and
\begin{align}\label{dhl1}
H_{ww}^{(c)}&=-\sigma\,v_c^{-1/2}\,(3f_{vv}^{(c)}+4\,v_c\,f_{vvv}^{(c)}-4\sigma\,\sqrt{v_c}\,f_{uvv}^{(c)}).
\end{align}

\item If $l\geq2$    then
$f_{uu}^{(c)}=f_{uv}^{(c)}=f_{vv}^{(c)}=0$ and
\begin{align}\label{dh2}
H_{ww}^{(c)}&=
\begin{cases}
v_c\,f_{uvv}^{(c)}\,(u_w^{(c)})^2+2\,v_c\,f_{vvv}^{(c)}u_w^{(c)}+f_{uvv}^{(c)},\quad & \mbox{for}\quad m=1,\\  \\
v_c\,f_{uvv}^{(c)}(u_w^{(c)})^2,\quad & \mbox{for}\quad m\geq2,
\end{cases}
\end{align}
where $u_w^{(c)}$ satisfies
\begin{equation}\label{du1}\begin{cases}
v_cf_{vvv}^{(c)}(u_w^{(c)})^2+2f_{uvv}^{(c)}u_w^{(c)}+f_{vvv}^{(c)}=0, & \mbox{for}\quad m=1,\\  \\
f_{vvv}^{(c)}\,u_w^{(c)}=0, & \mbox{for}\quad m\geq2.
\end{cases}
\end{equation}
\end{itemize}
\end{Lemma}
\vspace{0.3cm}
As a consequence we deduce the following conditions for critical points with $n=2$
\begin{pro}
Given $(x_c,\bt_c,u_c,v_c)\in \mC_{lm2}$ with  $v_c\neq0$, then it follows that
\begin{enumerate}
\item If $(x_c,\bt_c,u_c,v_c)\in\mC_{112}$, then
\begin{equation}\label{C112}
 3\sigma f_{vv}^{(c)}+4\sigma v_cf_{vvv}^{(c)}-4\sqrt{v_c}f_{uvv}^{(c)}\neq0.
\end{equation}
\item If $(x_c,\bt_c,u_c,v_c)\in\mC_{l12}$ with $l\geq2$ then
\begin{equation}\label{Cl12}
 f_{uvv}^{(c)^2}-v_cf_{vvv}^{(c)^2}\neq0.
\end{equation}
\item If $(x_c,\bt_c,u_c,v_c)\in\mC_{lm2}$ with $l,m\geq2$ then
\begin{equation}\label{Clm2}
f_{uvv}^{(c)}\neq0,\quad\mbox{and} \quad f_{vvv}^{(c)}=0.
\end{equation}
\end{enumerate}
\end{pro}

\begin{proof}
For $l=1$ we have that $m=1$ so that \eqref{C112} is a consequence of \eqref{dhl1}.  For $l\geq2$
and $m=1$ , by substituting the solution of \eqref{du1} into \eqref{dh2} we obtain the condition \eqref{Cl12}.

Finally suppose  that $l\geq 2,\,m \geq2$. Then according to \eqref{du1} and \eqref{dh2} we have that
$u_w^{(c)}\,f_{vvv}^{(c)}=0$, and
$H_{ww}^{(c)}=v_c(u_w^{(c)})^2\,f_{uvv}^{(c)}$. Therefore the
conditions for $H_{ww}^{(c)}\neq0$ are $f_{vvv}^{(c)}=0$ and
$f_{uvv}^{(c)}\neq 0$ which proves \eqref{Clm2}.

\end{proof}
Let us consider the system of  string equations \eqref{sis2} at points $(x,\bt,u,v)$ near a given critical point
 $(x_c,\bt_c,u_c,v_c)\in \mC_{lmn}$. In terms of symmetric variables it reads
\begin{equation}\everymath{\displaystyle}\label{sis2aa}
\begin{cases}
\oint_{\gamma}\dfrac{d z}{2\pi i\,z}\,
V_z(z,\bt)\,\V(z)\, =0,\\\\
\oint_{\gamma}\dfrac{d z}{2\pi i}\,V_z(z,\bt)\,\U(z)\,
=-2\,x.
\end{cases}
\end{equation}
One may introduce stretched variables not only for $x$ but also for $\bt$.  The  most
 symmetrical choice is
\begin{equation}\label{stt}
x=x_c+\overline{\epsilon}^{2n}\overline{x};\quad
t_j=t_{c,j}+\overline{\epsilon}^{2n}\overline{t}_j,\quad j\geq 1;\quad
\epsilon=\overline{\epsilon}^{2\,n+1}.
\end{equation}
Notice that $\epsilon\,\partial _{t_j}=\overline{\epsilon}\,\partial_{\overline{t}_j}$ so that \eqref{nev},
\eqref{stt}
and
\begin{equation}\label{seri}
u=u_c+\sum_{j\geq2}\overline{\epsilon}^j\overline{u}^{(j)},\quad
v=v_c+\sum_{j\geq2}\overline{\epsilon}^j\overline{v}^{(j)},
\end{equation}
are consistent with the Lax equations \eqref{tod}.

 We will next
concentrate on the case $n=2$ and will provide a recursive method
for determining expansions of the form \eqref{seri}  near
critical points in $\mC_{lm2}$. Thus  we set
$$x=x_c+\bar{\epsilon}^4\bar{x},\quad t_j=t_{c,j}+\bar{\epsilon}^4\bar{t}_j,\quad j\geq1, \quad
\bar{\epsilon}=\epsilon^{1/5}.$$
From \eqref{sis2aa}, equating the coefficients of order  $\overline{\epsilon}^j$ one finds
\begin{itemize}
\item For $0\leq j\leq 3$:
\begin{equation}\everymath{\displaystyle}\label{hod2}
\oint_{\gamma}\frac{dz}{2\pi i}V_z(z,\bt_c)\,\frac{\overline{V}^{(j)}}{z}=0,\quad
\oint_{\gamma}\frac{dz}{2\pi
i}V_z(z,\bt_c)\,\overline{U}^{(j)} =-2x_c\,\delta_{j0},
\end{equation}
\item For $j\geq4$:
\begin{equation}\everymath{\displaystyle}\begin{cases}\label{hod3}
\oint_{\gamma}\frac{dz}{2\pi i}V_z(z,\bt_c)\,\frac{\overline{V}^{(j)}}{z}+\oint_{\gamma}\frac{dz}{2\pi i}V_z(z,\overline{\bt})\,\frac{\overline{V}^{(j-4)}}{z}=0,\\  \\
\oint_{\gamma}\frac{dz}{2\pi
i}V_z(z,\bt_c)\, \overline{U}^{(j)}+\oint_{\gamma}\frac{dz}{2\pi i}V_z(z,\overline{\bt}) \,\overline{U}^{(j-4)}
=-2\overline{x}\,\delta_{j4}.
\end{cases}\end{equation}
\end{itemize}
Let us first analyze the system \eqref{hod2}. For $j=0$ it reduces
to the hodograph equations
\[
f_v^{(c)}=0,\quad f_u^{(c)}=2\,x_c,
\]
and
for $j=1$ is trivially verified since $\overline{U}^{(1)}\equiv \overline{V}^{(1)}\equiv 0$.
Furthermore, it is straightforward to see that for  $j=2$ both
equations in \eqref{hod2} reduce to
\begin{equation}\label{j2}
(\sigma\sqrt{v_c}\,\overline{u}^{(2)}+\overline{v}^{(2)})\,f_{vv}^{(c)}=0.\end{equation}
Hence to proceed further it is required  to distinguish the two types of critical points corresponding to  $\mC_{112}$ and  $\mC_{lm2}$ with $l\geq 2$.

\subsection*{Case $\mC_{112}$ }

In this case $f_{vv}^{(c)}\neq 0$ and \eqref{j2} implies
$$\overline{u}^{(2)}=\overline{w}^{(2)}=-\frac{\sigma}{\sqrt{v_c}}\overline{v}^{(2)}.$$
For  $j=3$ both equations in \eqref{hod2} lead to
$$\overline{w}^{(3)}=-\frac{\sigma}{\sqrt{v_c}}\overline{v}^{(3)}.$$
By setting  $j=4$ in \eqref{hod3}
 we find that
\begin{align*}
\overline{w}^{(4)}&=-\frac{\sigma}{\sqrt{v_c}}\left(\overline{v}^{(4)}+
\frac{1}{8}\overline{v}^{(2)}_{\overline{x}\overline{x}}+
\frac{1}{2\,v_c}\overline{v}^{(2)^2}\right)
-\frac{\sigma\,f_{v}(\overline{\bt},u_c,v_c)}{\sqrt{v_c}\,f_{vv}^{(c)}}
\\  \\             &+\dfrac{1}{6\,f_{vv}^{(c)}}\left[
f_{uvv}^{(c)}
-\sigma\,\sqrt{v_c}\,f_{vvv}^{(c)}\right]                \left(\overline{v}^{(2)}_{\overline{x}\overline{x}}
+\frac{6}{v_c}\overline{v}^{(2)^2}\right).
\end{align*}
Moreover $\overline{v}^{(2)}=-\sigma\,\sqrt{v_c}\,\overline{u}^{(2)}$ must satisfy the Painlev\'e I  equation
\begin{equation}\label{v2}
\Gamma^{(c)}\,
(\overline{v}^{(2)}_{\overline{x}\overline{x}}
+\frac{6}{v_c}\overline{v}^{(2)^2})=
6\,\Big(2\,\overline{x}-f_{u}(\overline{\bt},u_c,v_c)+\sigma\sqrt{v_c}\,f_{v}(\overline{\bt},u_c,v_c)\Big)
.
\end{equation}
where
\[
\Gamma^{(c)}:=-\sigma\,\frac{\sqrt{v_c}}{2}(3\,f_{vv}^{(c)}+4\,v_c\,f_{vvv}^{(c)}-4\,\sigma\,\sqrt{v_c}\,f_{uvv}^{(c)})
.
\]
Notice that since $(x_c,\bt_c,u_c,v_c)\in \mC_{112}$ then  Proposition 1 implies $\Gamma^{(c)}\neq 0$.

To proceed further one uses induction in the recurrence system for $\overline{U}^{(j)}$, $\overline{V}^{(j)}$
to prove that the only coefficients in the expansions of $\overline{U}^{(j)}$, $\overline{V}^{(j)}$ depending on $\overline{v}^{(j)}$,
$\overline{v}^{(j-1)}$, $\overline{v}^{(j-2)}$, $\overline{w}^{(j)}$,
$\overline{w}^{(j-1)}$ and $\overline{w}^{(j-2)}$ are those corresponding to $l=1$ or $2$. Moreover one finds
$$\everymath{\displaystyle}\begin{array}{lll}
\alpha^{(j)}_1&=&2\overline{v}^{(j)}+A^{(j)}_1,\\  \\
\alpha^{(j)}_2&=&2v_c\overline{v}^{(j-2)}_{\overline{x}\overline{x}}+12\overline{v}^{(2)}\overline{v}^{(j-2)}+
12v_c\overline{w}^{(2)}\overline{w}^{(j-2)}+A^{(j)}_2,\\  \\
\beta^{(j)}_1&=&4v_c\overline{w}^{(j)}+
\frac{v_c}{2}\overline{w}^{(j-2)}_{\overline{x}\overline{x}}+
4\overline{v}^{(2)}\overline{w}^{(j-2)}+4\overline{v}^{(j-2)}\overline{w}^{(2)}+
B^{(j)}_1,\\  \\
\beta^{(j)}_2&=&4v_c^2\overline{w}^{(j-2)}_{\overline{x}\overline{x}}
+24v_c(\overline{v}^{(2)}\overline{w}^{(j-2)}+\overline{v}^{(j-2)}\overline{w}^{(2)})
+B^{(j)}_2,\\  \\
\gamma^{(j)}_1&=&\overline{w}^{(j)}+C^{(j)}_1,\\  \\
\gamma^{(j)}_2&=&v_c\overline{w}^{(j-2)}_{\overline{x}\overline{x}}+6\overline{v}^{(2)}\overline{w}^{(j-2)}+
6\overline{v}^{(j-2)}\overline{w}^{(2)}+C^{(j)}_2,\\  \\
\eta^{(j)}_1&=&2\overline{v}^{(j)}+\frac{1}{4}\overline{v}^{(j-2)}_{\overline{x}\overline{x}}+2\overline{w}^{(2)}\overline{w}^{(j-2)}
+D^{(j)}_1,\\  \\
\eta^{(j)}_2&=&2v_c\overline{v}^{(j-2)}_{\overline{x}\overline{x}}+12v_c\overline{w}^{(2)}\overline{w}^{(j-2)}
+12\overline{v}^{(2)}\overline{v}^{(j-2)}+D^{(j)}_2.
\end{array}$$
where $A^{(j)}_i$, $B^{(j)}_i$, $C^{(j)}_i$ and $D^{(j)}_i$, $j\geq5$, $i=1,2$ are differential polynomials in
$\overline{v}^{(2)},\dots,\overline{v}^{(j-3)},\overline{w}^{(2)},\dots,\overline{w}^{(j-3)}$.
Then, for $j\geq5$ substituting \eqref{ues} into the first equation in \eqref{hod3} yields  expressions
$$\overline{w}^{(j)}=-\frac{\sigma}{\sqrt{v_c}}\overline{v}^{(j)}+K_j(\overline{\bt},\overline{v}^{(2)},\dots,\overline{v}^{(j-2)}),$$
where $K_j$ are differential polynomials in
$\overline{v}^{(2)},\ldots,\overline{v}^{(j-2)}$. Moreover, both equations in
\eqref{hod3} imply that $\overline{v}^{(j-2)}$ must verify a second
order linear differential equation of the form
\begin{equation}\label{vjm2}
\Gamma^{(c)}\,(\overline{v}^{(j-2)}_{\overline{x}\overline{x}}+\frac{12}{v_c}\overline{v}^{(2)}\overline{v}^{(j-2)})=
H_{j-2}(\overline{\bt},\overline{v}^{(2)},\dots,\overline{v}^{(j-3)}),
\end{equation}
where $H_{j-2}$ are  differential polynomials in
$\overline{v}^{(2)}$,...,$\overline{v}^{(j-3)}$.  In this way, we have a scheme
for determining the coefficients $\overline{u}^{(j)}$,
$\overline{w}^{(j)}$ in \eqref{expu}.

We notice that if
$\{\overline{v}^{(l)},\;l\geq2\}$ is a solution of
\eqref{v2}-\eqref{vjm2} ($j\geq5$), then as a consequence of the symmetry transformation \eqref{sym}-\eqref{simw}
we have that
$\{(-1)^l\overline{v}^{(l)},\;l\geq2\}$ is also  a solution of these
equations. Consequently, the differential polynomials $H_{2l+1}$,
$l\geq1$, in the right hand side of \eqref{vjm2} are odd polynomials
in $\overline{v}^{(2j+1)}$, $1\leq j\leq l-1$  and their derivatives
(for example $H_3(\overline{\bt},\overline{v}^{(2)})=0$). Hence we
may set
$\overline{v}^{(2j+1)}\equiv0$ for all $j\geq1$. Analogously, since the equations \eqref{v2}-\eqref{vjm2} can be written in terms of
$\{\overline{w}^{(l)},\;l\geq2\}$, we may set
$\overline{w}^{(2j+1)}\equiv0$ for all $j\geq1$.

\subsection*{Case $\mC_{lm2},\; l\geq 2$ }

In this case the system  \eqref{hod2} is trivially satisfied. Moreover, for $j=4$ we get  from  \eqref{hod3} that
\begin{equation}\everymath{\displaystyle}\label{sys2}
\begin{cases}
\Delta^{(c)}\,(v_c\overline{v}^{(2)}
_{\overline{x}\overline{x}}+3\,\overline{v}^{(2)^2}+3\,v_c
\overline{w}^{(2)^2})=
6\,\Big(f_{uvv}^{(c)}\,(f_{u}(\overline{\bt},u_c,v_c)-2\,\bar{x}) -v_c\,f_{vvv}^{(c)}\,f_{v}(\overline{\bt},u_c,v_c)\Big) ,
\\  \\
\Delta^{(c)}\,(v_c\overline{w}^{(2)}_{\overline{x}\overline{x}}+6\overline{w}^{(2)}\overline{v}^{(2)})=
6\,\Big(f_{uvv}^{(c)}\,f_{v}(\overline{\bt},u_c,v_c)-
f_{vvv}^{(c)}\,(f_{u}(\overline{\bt},u_c,v_c)-2\,\bar{x}) \Big),
\end{cases}
\end{equation}
where   we are denoting
\[
\Delta^{(c)}:=v_c\,(f_{vvv}^{(c)})^2-(f_{uvv}^{(c)})^2.
\]
From Proposition 1 we have that  $\Delta^{(c)}\neq 0$ for critical points in $\mC_{lm2}$ with $l\geq 2$.

In terms of the variables $u^{(\pm)}=\overline{v}^{(2)}\pm\sqrt{v_c}\overline{u}^{(2)}$,
the system \eqref{sys2} decouples into the two Painlev\'e I equations
$$
(f_{uvv}^{(c)}\pm\sqrt{v_c}f_{vvv}^{(c)})(v_c\,u_{\pm,\overline{x}\overline{x}}+3\, (u_{\pm})^2)=
6\,(2\,\overline{x}-f_{u}(\overline{\bt},u_c,v_c)\mp\sqrt{v_c}f_{v}(\overline{\bt},u_c,v_c)).
$$
Thus the doubling property is satisfied in this case.

In general, for $j\geq5$ \eqref{hod3}  leads to a second order
linear system for $\overline{w}^{(j-2)}$, $\overline{v}^{(j-2)}$ of
the form:
\begin{align}\everymath{\displaystyle}\label{newsys}
\nonumber 2\,v_c\;(\overline{v}^{(j-2)}_{\overline{x}\overline{x}}+6\,\overline{w}^{(2)}\overline{w}^{(j-2)})+12\,\overline{v}^{(2)}\overline{v}^{(j-2)}&=
{M}_{j-2}(\overline{t},\overline{w}^{(2)},\dots,\overline{w}^{(j-3)},\overline{v}^{(2)},\dots,\overline{v}^{(j-3)}),\\  \\
\nonumber v_c\,\overline{w}^{(j-2)}_{\overline{x}\overline{x}}+6\,\overline{w}^{(2)}
\overline{v}^{(j-2)}+6\,\overline{v}^{(2)}\overline{w}^{(j-2)}&=
{N}_{j-2}(\overline{t},\overline{w}^{(2)},\dots,\overline{w}^{(j-3)},\overline{v}^{(2)},\dots,\overline{v}^{(j-3)}).
\end{align}
where $M_{j-2}$ and $N_{j-2}$ are differential polynomials in
$\overline{w}^{(2)},\ldots, \overline{w}^{(j-3)}$ and
$\overline{v}^{(2)},\ldots,\overline{v}^{(j-3)}$. Thus, we have a
recursive procedure to construct  the coefficients
$\overline{w}^{(j)}$, $\overline{v}^{(j)}$, $j\geq2$.

Again, due to the symmetry \eqref{sym}-\eqref{simw}, it is clear that if $(\overline{w}^{(j)},\,\overline{v}^{(j)}),\,(j\geq2)$ is
solution of \eqref{newsys} then
$((-1)^j\overline{w}^{(j)},\,(-1)^j\overline{v}^{(j)}),\,(j\geq2)$
is also a solution of \eqref{newsys}. Hence for odd $j$  the differential
polynomials ${M}_{j-2}$, ${N}_{j-2}$  are
odd polynomials in $\overline{w}^{(2l+1)}$, $\overline{v}^{(2l+1)}$,
$l\geq1$ and their $\overline{x}$-derivatives. Therefore, we may set $\overline{w}^{(2l+1)}=0$,
$\overline{v}^{(2l+1)}=0$ for all $l\geq1$.

\subsection{Critical processes in ideal Hele-Shaw flows}

In view of the properties of asymptotic solutions of the KdV
 equation \cite{gra1}-\cite{gra3} and the NLS equation \cite{dub} it should be expected that  the inner expansions
provided by the double scaling method will be relevant when the solutions of the dispersionless or the dispersionful Toda hierarchies
  reach a  point of gradient catastrophe. We next consider an application to an ideal model of Hele-Shaw flows
  supplied by the Toda hierarchy.

A Hele-Shaw cell is a narrow gap between two plates filled with two fluids: say oil surrounding one or several bubbles of air. In  the set-up considered in \cite{lee3} (see also \cite{lee1}-\cite{lee2}) air is injected in two
fixed points of a simply-connected air bubble
 making  the bubble  break into two emergent bubbles.  Before the break-off the interface oil-air
 remains free of cusp-like singularities and develops a smooth neck. The reversed evolution describes the merging of two bubbles.
The analysis of \cite{lee3} concludes that before the merging, the
local structure of a small part of the interface containing the tips
of the bubbles is described
by a curve $Y=Y(X)$ which falls into universal classes characterized by two
even integers $(4\,n, 2),\, n\geq 1,$ and a finite number $2n$ of
deformation parameters $t_k$. Assuming symmetry of the curve
with respect to the $X$-axis, the general solution for the curve  in the $(4\,n, 2)$ class is
\begin{equation}\label{den}
Y(z):=\Big(\dfrac{\sum_{k=1}^{2n} (k+1)\,t_{k+1}\,z^k}{\sqrt{(z-a)(z-b)}}\Big)_\oplus\,\sqrt{(z-a)(z-b)},\quad X=z.
\end{equation}
where $a$ and $b$ are the positions of the bubbles tips.
 Due to the physical assumptions of the problem, the expansion
\begin{equation}\label{hoh}
Y(z)=\sum_{k=1}^{2n}
(k+1)\,t_{k+1}\,z^{k}+\sum_{k=0}^{\infty}\dfrac{Y_n}{z^n},\quad
z\rightarrow\infty,
\end{equation}
must satisfy the conditions $Y_0=t$ (physical
time) and $Y_1=0$ (bubble merging condition) which determine the
positions $a$, $b$ of the tips. However if $Y_1\neq 0$ the evolution
process leads to a critical point in which cusp-like singularities
appear.

As it was shown in \cite{lee3} the positions of the bubbles tips are determined by the pair of hodograph equations
\begin{equation}\label{ho}
\sum_{k=1}^\infty k\,t_k
r_{k-1}(u,v)=0,\quad
\sum_{k=1}^\infty
k\,t_k\,r_k(u,v)+2\,x=0,
\end{equation}
where $t_k=0$ for $k>2n$, $Y_1=2\,x$ and
\begin{equation}\label{d7b}
r :=\dfrac{z}{\sqrt{(z-u)^2-4v}}=\sum_{k\geq
0}\dfrac{r_k(u,v)}{z^k},\quad
a:=u-2\,\sqrt{v},\quad b:=u+2\,\sqrt{v}.
\end{equation}
These are precisely the hodograph equations \eqref{hods}. Thus  the double scaling limit method can be used to
regularize the  solutions of \eqref{den}-\eqref{ho} at critical points in terms of inner expansions of solutions of
\eqref{sis2}.

As an example let us analyze the critical process of a merging of two bubbles studied in
section VII of \cite{lee3} . We set $t:=t_1$, $t_2=0$, $t_n=0$, $n>3$
and fix $t_3$ to a given constant value $c$.
Thus
the  Hele-Shaw interface is locally  characterized by the curve
\begin{equation}\label{inter}
Y(X)=3\,c\,(X+u)\sqrt{(X-u)^2-4v},
\end{equation}
and \eqref{ho} reduces to
\begin{equation}\label{hod0}
t+3\,c\,(u^2+2v)\,=\,0,\quad
6\,c\,v\,u+x\,=\,0.
\end{equation}
This is the hodograph system \eqref{crite} and its solution is given by
\begin{align}\label{soltoda}
\nonumber v=&\frac{t^2}{2^{2/3}\,9 \,c\,\sqrt[3]{9\, \sqrt{81\,c^2\, x^4 +4 \,c\, x^2\,
   t^3}-81 \,c\, x^2 -2 \,t^3}}-\frac{t}{18 \,c}
\\\\  \nonumber &+\frac{\sqrt[3]{9\, \sqrt{81\, c^2\,x^4+4 c\,t^3 x^2 }-81 \,c\, x^2 -2 t^3}}
   {18 \,\sqrt[3]{2}\, c},\quad u=-\frac{x}{6\,c\,v}.
\end{align}
As we have seen above the  critical points $(x_c,t_c,t_{c3},u_c,v_c)$ with
$t_c\neq 0$  are in $\mC_{112}$ so that our scheme can be
applied to provide a dispersive regularization of \eqref{soltoda}
near critical points.

Setting $x=x_c$ in \eqref{soltoda} one finds that the outer approximations for $u$ and $v$ near $t=t_c$ are given by
\begin{equation}\label{ap1}
u\,\sim\,u_{out}(t):=u_c\,-\,\frac{1}{3}\sqrt{\frac{1}{c}(t_c-t)},\quad
v\,\sim\,v_{out}(t):=v_c\,+\,\frac{u_c}{3}\sqrt{\frac{1}{c}(t_c-t)}, \quad
\mbox{as} \quad t\rightarrow t_c^-,
\end{equation}
where $u_c=\sigma\,\sqrt{v_c}$.
Since we are considering a critical point in $\mC_{112}$ we introduce the stretched variables
\[
x=x_c+\overline{\epsilon}^4\,\overline{x},\quad t=t_c+\overline{\epsilon}^4\,\overline{t}
\]
and  consider the inner expansions for $u$ and $v$ \be\label{p1}
u\sim u_{in}:=u_c+\overline{\epsilon}^2\,\overline{u}^{(2)}(\overline{x}-u_c\,\overline{t}),\qquad
v\sim  v_{in}(\overline{t}): =v_c+\overline{\epsilon}^2\,\overline{v}^{(2)}(\overline{x}-u_c\,\overline{t}),\quad
\mbox{as} \quad \overline{t}\rightarrow 0, \ee
where $\overline{u}^{(2)}=-\overline{v}^{(2)}/u_c$  and $\overline{v}^{(2)}$  verifies the Painlev\'e I equation
\begin{equation}\label{eqV2}
\overline{v}^{(2)}_{\overline{x}\overline{x}}+\frac{6}{u_c^2}(\overline{v}^{(2)})^2\,=\,\frac{2}{3\,c\,u_c}(\overline{x}-u_c\,\overline{t}).
\end{equation}

 The inner approximation at $\overline{x}=0$ must match the outer approximation  in an overlap interval which has both $t-t_c$ small and  $\overline{t}$ large. Writing the outer approximations \eqref{ap1} in terms of the inner variable $\overline{t}$ we have
\[
u_{out}(t)=u_c-\overline{\epsilon}^2\,\frac{1}{3}\sqrt{-\frac{\overline{t}}{c}},\quad
v_{out}(t)=v_c+\overline{\epsilon}^2\,\frac{u_c}{3}\sqrt{-\frac{\overline{t}}{c}}.
\]
Hence, it is clear that matching requires a solution $\overline{v}^{(2)}(\overline{x}-u_c\,\overline{t})$ of \eqref{eqV2} satisfying
\begin{equation}\label{asym2}
\overline{v}^{(2)}(-u_c\,\overline{t})\,\sim\,\frac{u_c}{3}\sqrt{-\frac{\overline{t}}{c}},\quad
\mbox{as} \quad \overline{t}\rightarrow-\infty.
\end{equation}
Now, if we introduce the change of variables
\[
W\,=\,-\left(\frac{3\,c}{2\,u_c^2}\right)^{2/5}\overline{v}^{(2)},\quad
\xi\,=\left(\frac{2}{3\,c\,u_c^3}\right)^{1/5}\,(\overline{x}-u_c\,\overline{t}),
\]
it follows that $W$ must satisfy the P-I equation
\[
W_{\xi\xi}=6\,W^2-\,\xi,
\]
 and
$$
W\sim-\sqrt{\frac{\xi}{6}},\quad \mbox{as}\quad
\xi\rightarrow \infty,
$$
so that $W$ must be the tritronqu\'ee
solution of the P-I equation \cite{bout}-\cite{josi}. Using a numerical  approximation of $W$
one finds \cite{toda} that  the regularized evolution is as follows: the
right bubble develops a cusp, then a new bubble appears at this cusp and it
grows until it merges with the tip of the left bubble.
Finally, the tip of the left bubble absorbs the new one and stays
joined to the right bubble.

\vspace{0.5cm}

\noindent {\large{\bf Acknowledgements}}

\vspace{0.3cm}

The authors  wish to thank the  Spanish Ministerio de Educaci\'on y Ciencia (research project FIS2005-00319) and
 the European Science Foundation (MISGAM programme) for their  support.

\end{document}